# Mathematics of CLIFFORD - A Maple Package for Clifford and Graßmann Algebras

Rafał Abłamowicz[1)]    Bertfried Fauser[2)]

February 17, 2005

**Abstract**

CLIFFORD performs various computations in Graßmann and Clifford algebras. It can compute with quaternions, octonions, and matrices with entries in $C\ell(B)$ - the Clifford algebra of a vector space $V$ endowed with an arbitrary bilinear form $B$. Two user-selectable algorithms for the Clifford product are implemented: `cmulNUM` - based on Chevalley's recursive formula, and `cmulRS` - based on a non-recursive Rota-Stein sausage. Graßmann and Clifford bases can be used. Properties of reversion in undotted and dotted wedge bases are discussed.

## 1 Introduction

As many programs CLIFFORD emerged from a practical problem. Relatively complicated algebraic manipulations with octonions, which can be performed in **spin**(7), started a thread which has now developed into a multi purpose algebra tool. It is the basic structure of a vector space $V$ endowed with a quadratic form $Q$ which is common to a vast host of mathematical, physical and engineering problems, and which allows one to build naturally –i.e., in a categorial sense 'for free'– an algebra structure, the Clifford algebra $C\ell(V,Q)$.

---

1) DEPARTMENT OF MATHEMATICS, BOX 5054, TENNESSEE TECHNOLOGICAL UNIVERSITY, COOKEVILLE, TN 38505, USA
rablamowicz@tntech.edu,
http://math.tntech.edu/rafal/

2) MAX PLANCK INSTITUT FÜR MATHEMATIK, INSELSTRASSE 22-26, D-04103 LEIPZIG, GERMANY,
Bertfried.Fauser@uni-konstanz.de,
http://clifford.physik.uni-konstanz.de/~ fauser/

While in a conventional vector calculus one makes a good use of the vector space structure, one does not have yet a vector algebra since vector multiplication is missing. Having established a Clifford algebra structure provides one with an entirely new formalism that now can be applied to solving completely different problems.

In this sense, `CLIFFORD` is a basic tool for all such investigations and applications which can be carried in finite dimensional vector spaces equipped with a quadratic form or, equivalently, with a symmetric bilinear form commonly referred to as an *inner* or a *scalar* product. The intrinsic abilities of Maple even allow one to use `CLIFFORD` in projective and affine geometries while visualizing complicated incidence relations that is helpful, e.g., for image processing, visual perception and robotics.

The authors of `CLIFFORD` have been interested in fundamental questions about $q$-deformed symmetries and quantum field theory. A reasonable number of new results has been derived by using systematically the ability to *compute* with a Computer Algebra System (CAS) at hand. Moreover, just asking questions such as "What is the most general element fulfilling ...?" has led to unexpected results and new insights. Testing of theorems, usually to check the consistency of the software, has led from time to time to counter examples that have made a rethinking and a more careful restatement of those theorems necessary. However, the most striking ability of `CLIFFORD` is that it is unique in being able to handle Clifford algebras of an *arbitrary bilinear form* not restricted by symmetry and not directly related to any quadratic form. Since it is now well known that such structures are related to Hopf algebraic twists, later versions of `CLIFFORD` make an extensive use of a process called *Rota-Stein cliffordization*, which turns out to be a Drinfeld twist of a Graßmann Hopf algebra [9].

The present paper introduces the reader to the package. It is assumed that she is already familiar with Maple [22], a general purpose CAS; if not please consult e.g., [23]. Of course, such a paper cannot be a *user guide* but may only be a demonstration of the usability and strength of the package. The interested reader is invited to download the package and take a closer look at the online documentation in the Maple help browser or to download a pdf file with over 500 help pages. Therein we provide also further mathematical background and references and a detailed description for every function. However, the present article provides also a quick –and dirty– introduction which is sufficient to get started.

A list of goals behind an *experimental mathematics*, which are the guiding beacons for this work, is given in [5] that describes the supplementary package `BIGEBRA` for computations with tensor and Hopf algebras.

## 2  Notations and basic computations

`CLIFFORD` uses as default a standard Graßmann basis (Graßmann multivectors) in $\bigwedge V$ where $V = \text{span}\{\mathbf{e}_i \,|\, 1 \leq i \leq n\}$ for $1 \leq n \leq 9$. Then $\bigwedge V = \text{span}\{\mathbf{e}_i \wedge \mathbf{e}_j \wedge \ldots \wedge \mathbf{e}_k \,|\, 0 \leq i < j < \ldots < k \leq n\}$. In `CLIFFORD` these basis monomials

are written as strings {*Id, e1, ..., e9, e1we2, e1we3, ..., e1we2we3, ...*} although they can be aliased to shorten input. Here *e1we2* is a string that denotes $\mathbf{e}_1 \wedge \mathbf{e}_2$ and Id denotes the identity **1** in $\bigwedge V$. However, CLIFFORD can also use one-character long symbolic indices as in *eiwej*. Thus, in principle, it can compute with Clifford algebras in dimensions higher than 9. For example, when $n = 3$, Graßmann basis monomials are:
> `W=cbasis(3);`

$$W = [\mathit{Id},\, \mathit{e1},\, \mathit{e2},\, \mathit{e3},\, \mathit{e1we2},\, \mathit{e1we3},\, \mathit{e2we3},\, \mathit{e1we2we3}]$$

but aliases can also be used to shorten input/output:
> `eval(makealiases(3));`

$$I,\, \mathit{e12},\, \mathit{e21},\, \mathit{e13},\, \mathit{e31},\, \mathit{e23},\, \mathit{e32},\, \mathit{e123},\, \mathit{e132},\, \mathit{e213},\, \mathit{e231},\, \mathit{e312},\, \mathit{e321}$$

In the above, $\mathit{eijk} = \mathit{eiwejwek}$ is the wedge product of three 1-vectors: $\mathbf{e}_i, \mathbf{e}_j, \mathbf{e}_k$. Thus, the most general element in the Graßmann algebra $\bigwedge V$ is a Graßmann polynomial which is just a linear combination of Graßmann basis monomials with real coefficients. Notice that symbolic indices are allowed:
> `p1:=Id+4.5*ei-alpha*e1we2we3;`

$$p1 := \mathit{Id} + 4.5\, \mathit{ei} - \alpha\, \mathit{e123}$$

Reordering of Graßmann monomials can be explicitly accomplished with a procedure `reorder`. CLIFFORD procedures ordinarily return their results in the standard (reordered) basis.
> `p2:=-e3we2we1-x0*Id+x12*e2we1+a*ejwei; reorder(p2);`

$$p2 := -\mathit{e321} - x0\,\mathit{Id} + x12\,\mathit{e21} + a\,\mathit{ejwei}$$
$$\mathit{e123} - x0\,\mathit{Id} - x12\,\mathit{e12} - a\,\mathit{eiwej}$$

The wedge product $\wedge$ is computed with a procedure `wedge` or its ampersand counterpart $\&w$:
> `wedge(e1,e2),e1 &w e2; wedge(ea,eb,ec),ea &w eb &w ec; p1 &w p2;`

$$\mathit{e12},\, \mathit{e12}$$
$$\mathit{eawebwec},\, \mathit{eawebwec}$$
$$\mathit{e123} - x0\,\mathit{Id} - 4.500000000\,x0\,\mathit{e1} + \alpha\,x0\,\mathit{e123} - x12\,\mathit{e12}$$

Following Chevalley's recursive definition, a Clifford product can be introduced in $\bigwedge V$ by means of a left $\lrcorner_B$ (or right $\llcorner_B$) contraction dependent on an arbitrary bilinear form $B : V \times V \to \mathbb{R}$ (see Sect. 3). This leads to elements of the Clifford algebra $C\ell(B)$ expanded into multivectors and makes the Clifford multiplication implicitly dependent on $B$. The associative Clifford product is given by a procedure `cmul` or its infix form $\&c$.
> `cmul(e1,e2),&c(e1,e2); cmul(ea,eb,ec);`

$$\mathit{e12} + B_{1,2}\,\mathit{Id},\; \mathit{e12} + B_{1,2}\,\mathit{Id}$$
$$\mathit{eawebwec} + B_{b,c}\,\mathit{ea} - B_{a,c}\,\mathit{eb} + B_{a,b}\,\mathit{ec}$$

Computations in $C\ell(K)$ and $C\ell(B)$ can be performed in the same worksheet since the name of a bilinear form can be passed to `cmul` as a parameter. For example,
> `cmul[K](e1,e2),&c[K](e1,e2); cmul[K](ei,ej,ek);`

$$e12 + K_{1,2}\, Id, e12 + K_{1,2}\, Id$$
$$eiwejwek + K_{j,k}\, ei - K_{i,k}\, ej + K_{i,j}\, ek$$

Of course, the forms $B$ and $K$ can be numeric or symbolic. For example, when
> `B:=matrix(2,2,[1,a,a,1]);`

$$B := \begin{bmatrix} 1 & a \\ a & 1 \end{bmatrix}$$

then the Graßmann basis for $C\ell(B)$ or $\bigwedge V$ will be:
> `cbas:=cbasis(2);`

$$cbas := [Id,\ e1,\ e2,\ e12]$$

while the Clifford multiplication table of the basis Graßmann monomials will look as follows:
> `MultTable:=matrix(4,4,(i,j)->cmul(cbas[i],cbas[j]));`

$$MultTable := \begin{bmatrix} Id & e1 & e2 & e12 \\ e1 & Id & e12 + a\, Id & e2 - a\, e1 \\ e2 & -e12 + a\, Id & Id & a\, e2 - e1 \\ e12 & a\, e1 - e2 & e1 - a\, e2 & (-1 + a^2)\, Id \end{bmatrix}$$

Of course, irrespective of the bilinear form chosen, the Graßmann multiplication table will always remain as:
> `wedgetable:=matrix(4,4,(i,j)->wedge(cbas[i],cbas[j]));`

$$wedgetable := \begin{bmatrix} Id & e1 & e2 & e12 \\ e1 & 0 & e12 & 0 \\ e2 & -e12 & 0 & 0 \\ e12 & 0 & 0 & 0 \end{bmatrix}$$

Let $B = g + F$ where $g$ and $F$ are the symmetric and the antisymmetric parts of $B$:
> `g,F:=matrix(2,2,[g11,g12,g12,g22]),matrix(2,2,[0,F12,-F12,0]);`
> `B:=evalm(g+F);`

$$g, F := \begin{bmatrix} g11 & g12 \\ g12 & g22 \end{bmatrix},\ \begin{bmatrix} 0 & F12 \\ -F12 & 0 \end{bmatrix}$$

$$B := \begin{bmatrix} g11 & g12 + F12 \\ g12 - F12 & g22 \end{bmatrix}$$

Then, the Clifford multiplication table of the basis monomials in $C\ell(B)$ will be as follows:

```
> MultTable:=matrix(4,4,(i,j)->cmul(cbas[i],cbas[j]));
```

$MultTable :=$
$[Id, e1, e2, e12]$
$[e1, g11\,Id, e12 + (g12 + F12)\,Id, g11\,e2 - (g12 + F12)\,e1]$
$[e2, (g12 - F12)\,Id - e12, g22\,Id, (g12 - F12)\,e2 - g22\,e1]$
$[e12, (g12 - F12)\,e1 - g11\,e2, g22\,e1 - (g12 + F12)\,e2,$
$(g12^2 - F12^2 - g22\,g11)\,Id - 2\,e12\,F12]$

Observe, that the "standard" anticommutation relations

$$\mathbf{e}_i \mathbf{e}_j + \mathbf{e}_j \mathbf{e}_i = (B_{i,j} + B_{j,i})\mathbf{1} = 2g_{i,j}\mathbf{1}$$

are satisfied by the generators $\mathbf{e}_i$, $i = 1, 2, \ldots, n$, irrespective of the presence of the antisymmetric part $F$ in $B$. For example,

```
> cmul[g](e1,e2)+cmul[g](e2,e1);
```

$$2\,Id\,g12$$

```
> cmul[B](e1,e2)+cmul[B](e2,e1);
```

$$(g12 + F12)\,Id + (g12 - F12)\,Id$$

```
> clisort(simplify(%));
```

$$2\,g12\,Id$$

It is well known [16, 19] that real Clifford algebras $C\ell(V, Q) = C\ell_{p,q}$ are classified in terms of the signature $(p, q)$ of $Q$ and the dimension $\dim V = n = p + q$. Information about all Clifford algebras $C\ell_{p,q}$, $1 \leq n \leq 9$, for any signature $(p, q)$ has been pre-computed and stored in CLIFFORD, and it can be retrieved with a procedure clidata. For example, for the Clifford algebra $C\ell_{2,0}$ (also denoted as $C\ell_2$) of the Euclidean plane $\mathbb{R}^2$ we find:

```
> clidata([2,0]); #Clifford algebra of the Euclidean plane
```

$$[real, 2, simple, \frac{1}{2}Id + \frac{1}{2}e1, [Id, e2], [Id], [Id, e2]]$$

The meaning of the first three entries in the above output list is that $C\ell_2$ is a simple algebra isomorphic to $\text{Mat}(2, \mathbb{R})$. The 4th entry in the list gives a primitive idempotent $f$ that has been used to generate a minimal left spinor ideal $S = C\ell_2 f$ and, subsequently, the left spinor (lowest dimensional and faithful) representation of $C\ell_2$ in $S$. In general it is known that, depending on $(p, q)$ and $n = \dim V$, the spinor ideal $S = C\ell_{p,q} f$ is a right $K$-module where $K$ is either $\mathbb{R}, \mathbb{C}$, or $\mathbb{H}$ for simple Clifford algebras when $(p - q) \neq 1 \mod 4$, or $\mathbb{R} \oplus \mathbb{R}$ and $\mathbb{H} \oplus \mathbb{H}$ for semisimple algebras when $(p - q) = 1 \mod 4$ [14, 17]. Elements in the 5th entry (here $[Id, e2]$) generate a real basis in $S$ with respect to $f$, that is, $S = \text{span}\{Id \,\&c\, f, e2 \,\&c\, f\} = \text{span}\{f, e2 \,\&c\, f\}$. Elements in the 6th entry span a subalgebra $F$ of $C\ell(Q)$ that is isomorphic to $K$. In the case of $C\ell_2$ we find that $F = \text{span}\{Id\} \cong \mathbb{R}$. The last entry in the output gives $2^k$ generators of $S$ (with respect to $f$) viewed as a right module over $K$ where $k = q - r_{q-p}$ and $r$ is the Radon-Hurwitz number.[1] Number $k$ is the number of factors $\frac{1}{2}(\mathbf{1} + T_i)$, where $\{T_i\}$, $i = 1, \ldots, k$, is a set of commuting basis Graßmann monomials squaring in $C\ell(Q)$ to $\mathbf{1}$, whose product gives a primitive idempotent $f$ in $C\ell(Q)$. Spinor representation for all Clifford algebras $C\ell(Q)$, $1 \le n = p + q \le 9$, and for any signature $(p, q)$ has been pre-computed [1] and can be retrieved from CLIFFORD with a procedure matKrepr. For example, 1-vectors $\mathbf{e}_1$ and $\mathbf{e}_2$ in $C\ell_2$ have the following spinor representation in the basis $\{f, e2 \,\&c\, f\}$ of $S = C\ell_2 f$:[2]

```
> matKrepr([2,0]);
```

$$[e1 = \begin{bmatrix} 1 & 0 \\ 0 & -1 \end{bmatrix}, e2 = \begin{bmatrix} 0 & 1 \\ 1 & 0 \end{bmatrix}]$$

In another example, Clifford algebra $C\ell_3$ of $\mathbb{R}^3$ is isomorphic with $\text{Mat}(2, \mathbb{C})$:

```
> B:=linalg[diag](1,1,1):clidata([3,0]);
```

$$[complex, 2, simple, \frac{1}{2}Id + \frac{1}{2}e1, [Id, e2, e3, e23], [Id, e23], [Id, e2]]$$

and its spinor representation is given in terms of Pauli matrices:

```
> matKrepr([3,0]);
```

$$[e1 = \begin{bmatrix} 1 & 0 \\ 0 & -1 \end{bmatrix}, e2 = \begin{bmatrix} 0 & 1 \\ 1 & 0 \end{bmatrix}, e3 = \begin{bmatrix} 0 & -e23 \\ e23 & 0 \end{bmatrix}]$$

Notice that $F = \text{span}\{Id, e23\}$ ($e23 = e2we3$) is a subalgebra of $C\ell_3$ isomorphic to $\mathbb{C}$. Since Pauli matrices belong to $\text{Mat}(2, F)$, it is necessary for CLIFFORD to compute with Clifford matrices, that is, matrices of a type climatrix with entries in a Clifford algebra.

```
> M1,M2,M3:=rhs(%[1]),rhs(%[2]),rhs(%[3]);
```

$$M1, M2, M3 := \begin{bmatrix} 1 & 0 \\ 0 & -1 \end{bmatrix}, \begin{bmatrix} 0 & 1 \\ 1 & 0 \end{bmatrix}, \begin{bmatrix} 0 & -e23 \\ e23 & 0 \end{bmatrix}.$$

Of course Pauli matrices satisfy the same defining relations as the basis vectors $\mathbf{e}_1, \mathbf{e}_2,$ and $\mathbf{e}_3$:[3] For example:

```
> 'M1 &cm M2 + M2 &cm M1' = evalm(M1 &cm M2 + M2 &cm M1);
> 'e1 &c e2 + e2 &c e1'=e1 &c e2 + e2 &c e1;
```

---

[1] Type ?RHnumber in a Maple session when CLIFFORD is installed for more help.

[2] We use the sloppy notation $1 \equiv \mathbf{1}$ in Clifford algebra valued matrices which produces a simpler display.

[3] Here $\&cm$ is a matrix product where Clifford multiplication is applied to the matrix entries. See ?&cm for more information.

$$M1 \ \&cm \ M2 \ + \ M2 \ \&cm \ M1 = \begin{bmatrix} 0 & 0 \\ 0 & 0 \end{bmatrix}$$

$$e1 \ \&c \ e2 \ + \ e2 \ \&c \ e1 = 0$$

```
> 'M1 &cm M1' = evalm(M1 &cm M1),'M2 &cm M2' = evalm(M2 &cm M2),
> 'M3 &cm M3' = evalm(M3 &cm M3);
> 'e1 &c e1' = e1 &c e1,'e2 &c e2' = e2 &c e2,'e3 &c e3' = e3 &c e3;
```

$$M1 \ \&cm \ M1 = \begin{bmatrix} 1 & 0 \\ 0 & 1 \end{bmatrix}, \ M2 \ \&cm \ M2 = \begin{bmatrix} 1 & 0 \\ 0 & 1 \end{bmatrix}, \ M3 \ \&cm \ M3 = \begin{bmatrix} 1 & 0 \\ 0 & 1 \end{bmatrix}$$

$$e1 \ \&c \ e1 = Id, \ e2 \ \&c \ e2 = Id, \ e3 \ \&c \ e3 = Id$$

The procedure `matKrepr` gives the linear isomorphism $C\ell(Q) \simeq \mathrm{Mat}(2, \mathbb{R})$, and, in general, $C\ell(Q) \simeq \mathrm{Mat}(2^k, K)$, where $K = \mathbb{R}, \mathbb{C}, \mathbb{H}$, for simple algebras and $C\ell(Q) \simeq \mathrm{Mat}(2^k, K) \oplus \mathrm{Mat}(2^k, K)$, where $K = \mathbb{R}, \mathbb{H}$, for semisimple algebras. In this latter case, it is customary to represent an element in $C\ell(Q)$ in terms of a single matrix over a double field $\mathbb{R} \oplus \mathbb{R}$ or $\mathbb{H} \oplus \mathbb{H}$ rather than as pair of matrices.[4]

One can easily list signatures of the quadratic form $Q$ for which $C\ell(Q)$ is simple or semisimple. For more information, type `?all_sigs`. For example, $C\ell_{1,3}$ has a spinor representation given in terms of 2 by 2 quaternionic matrices whose entries belong to a subalgebra $F$ of $C\ell_{1,3}$ spanned by $\{Id, e2, e3, e2we3\}$ :

```
> B:=linalg[diag](1,-1,-1,-1):clidata([1,3]);
```

$[quaternionic, \ 2, \ simple, \ \frac{1}{2} Id + \frac{1}{2} e1we4, \ [Id, e1, e2, e3, e12, e13, e23, e123],$
$[Id, e2, e3, e23], [Id, e1]]$

```
> matKrepr([1,3]); #quaternionic matrices
```

$$[e1 = \begin{bmatrix} 0 & 1 \\ 1 & 0 \end{bmatrix}, \ e2 = \begin{bmatrix} e2 & 0 \\ 0 & -e2 \end{bmatrix}, \ e3 = \begin{bmatrix} e3 & 0 \\ 0 & -e3 \end{bmatrix}, \ e4 = \begin{bmatrix} 0 & -1 \\ 1 & 0 \end{bmatrix}]$$

CLIFFORD includes several special-purpose procedures to deal with quaternions and octonions (type `?quaternion` and `?octonion` for help). In particular, following [18], octonions are treated as para-vectors in $C\ell_{0,7}$ while their non-associative multiplication, defined via Fano triples, is related to the Fano projective plane $\mathbb{F}_2$ (see `?omultable`, or `?Fano_triples` for more information). Since the bilinear form $B$ can be degenerate[5], one can use CLIFFORD to perform computations in Clifford algebras $C\ell_{p,q,d}$ of degenerate quadratic form $Q$ of signature $(p, q)$ and the dimension of the radical $d$. For example, the Clifford algebra $C\ell_{0,3,1}$ of the quadratic form $Q(\mathbf{x}) = -x_1^2 - x_2^2 - x_3^2$ where $\mathbf{x} = x_1\mathbf{e}_1 + x_2\mathbf{e}_2 + x_3\mathbf{e}_3 + x_4\mathbf{e}_4 \in \mathbb{R}^4$ is used in robotics to represent rigid motions in $\mathbb{R}^3$ and screw motions in terms of dual quaternions [21].

---

[4]Procedures `adfmatrix` and `mdfmatrix` add and multiply matrices of type `dfmatrix` over such double fields. For more information see `?matKrepr`.

[5]When $B \equiv 0$ then $C\ell(V, B) = C\ell_{0,0,n} = \bigwedge V$ and computations in the Graßmann algebra $\bigwedge V$ can then be done with CLIFFORD.

Thus, `CLIFFORD` is a repository of mathematical knowledge about Clifford algebras of a quadratic form in dimensions 1 through 9. Together with a supplementary package `BIGEBRA` [3] it can be extended to graded tensor products of Clifford algebras in higher dimensions. The `BIGEBRA` package is described in [5]. For more information about any `CLIFFORD` or `BIGEBRA` procedure, type `?Clifford` or `?Bigebra` to see its help page through the Maple browser. For a computation of spinor representations with `CLIFFORD` we refer to [1].

## 3 Clifford product

Since the Clifford product provides the main functionality of the `CLIFFORD` package, the best available mathematics has been used to program it. Two internal user-selectable functions `cmulRS` and `cmulNUM` encode the Clifford product but the user normally does not use either one. Instead, she uses a wrapper function $\&c[K](arg1, arg2, \ldots)$ or $\text{cmul}[K](arg1, arg2, \ldots)$ that passes the name of a bilinear form $K$ to either `cmulRS` or `cmulNUM`, whichever one has been selected. The wrapper function can also act on any number of arguments of type `clipolynom` (this makes sense since the Clifford product is associative) and on a much wider class of types including Clifford matrices of type `climatrix`. It can also accept Clifford polynomials in other bases such as the Clifford basis $\{\mathbf{1}, \mathbf{e}_i, \mathbf{e}_i \,\&\text{C}\, \mathbf{e}_j, \&\text{C}(\mathbf{e}_i, \mathbf{e}_j, \mathbf{e}_k), \ldots\}$ where &C denotes the unevaluated Clifford product. Clifford basis differs from the Graßmann exterior basis when $B$ is not a diagonal matrix.[6]

The two internal Clifford multiplication procedures are appropriate for different purposes. While `cmulNUM` is fast on sparse numeric matrices and on numeric matrices in general when $\dim V \geq 5$, procedure `cmulRS` was designed for symbolic calculations. Since `cmulRS` computes reasonably well in the numeric sparse case up to $\dim V = 5$, it was chosen as the default product of the package.[7] Both procedures take as input two Clifford monomials of type `clibasmon` along with a third argument $K$ of a type `name, symbol, matrix` or `array` which contains the chosen bilinear form for $C\ell(K)$.

Via a procedure `useproduct`, the user can select either `cmulRS` or `cmulNUM` depending which procedure is expected to give a better performance for the bilinear form in use. In addition, the user can also supply a new product function (not necessarily a Clifford product) acting on two basis monomials. The wrapper function uses then the selected algorithm to compute the Clifford product.

---

[6]Procedures converting between Graßmann and Clifford bases belong to a supplementary package `Cliplus` [3] while Clifford polynomials expressed in the Clifford basis are of type `cliprod`. Type `?cliprod` for more information.

[7]When `CLIFFORD` is loaded, it is setup to use `cmulRS` by default. Type `?useproduct` for help how to change to `cmulNUM`.

## 3.1 Recursive procedure `cmulNUM`

The evaluation of Clifford products in the Graßmann basis is quite involved and it is normally done by a recursive process that involves Chevalley deformation. This algorithm is encoded in `cmulNUM`. Unfortunately, during the recursive evaluation many superfluous terms appear that later cancel out at the next recursive call. When the bilinear form is sparse numeric, many branches of the recursion are cut out by Maple quite early due to automatic evaluation that takes precedence over the recursion. In this case, the superfluous terms disappear and are not passed on to the next recursive step. However, in the symbolic case, in general, all these terms might be non-zero which prevents fast completion of the recursion. Fortunately, Hopf combinatorial methods free of the drawbacks of the recursion can also be applied and have been encoded in `cmulRS`. Thus, the two ways to evaluate the Clifford product in `CLIFFORD` have emerged.

We introduce the Chevalley deformation and the Clifford map to explain the algorithm used in `cmulNUM`. The Clifford map $\gamma_\mathbf{x}$ is defined on $u \in \bigwedge V$ as

(i) $\gamma_\mathbf{x}(u) = \mathtt{LC}(x, u, B) + \mathtt{wedge}(x, u) = \mathbf{x} \,\lrcorner_B u + \mathbf{x} \wedge u$

(ii) $\gamma_\mathbf{x} \gamma_\mathbf{y} = \gamma_{\mathbf{x} \wedge \mathbf{y}} + B(\mathbf{x}, \mathbf{y}) \gamma_\mathbf{1}$

(iii) $\gamma_{a\mathbf{x}+b\mathbf{y}} = a\gamma_\mathbf{x} + b\gamma_\mathbf{y}$

where $\mathbf{x}, \mathbf{y} \in V$ (see, for example, [19]). One knows how to compute with the wedge $\mathbf{x} \wedge u$ and the left contraction $\mathbf{x} \,\lrcorner_B u$ with respect to the bilinear form $B$ (in `CLIFFORD`, the left contraction $\lrcorner_B$ is given by the procedure $\mathtt{LC}(\mathbf{x}, u, B)$). Following Chevalley, the left contraction has the following properties:

(i) $\mathbf{x} \,\lrcorner_B \mathbf{y} = B(\mathbf{x}, \mathbf{y})$

(ii) $\mathbf{x} \,\lrcorner_B (u \wedge v) = (\mathbf{x} \,\lrcorner_B u) \wedge v + \hat{u} \wedge (\mathbf{x} \,\lrcorner_B v)$

(iii) $(u \wedge v) \,\lrcorner_B w = u \,\lrcorner_B (v \,\lrcorner_B w)$

where $\mathbf{x} \in V$, $u, v \in \bigwedge V$ and $\hat{u}$ is the Graßmann grade involution. Hence we can use the Clifford map $\gamma_\mathbf{x}$ (Chevalley deformation of the Graßmann algebra) to define a Clifford product of a one-vector $\mathbf{x}$ and a multivector $u$ as

$$\mathbf{x} u = \mathbf{x} \,\lrcorner_B u + \mathbf{x} \wedge u.$$

Analogous formula can also be given for a right Clifford map using the right contraction $\llcorner_B$ implemented as the procedure `RC`.

The Clifford product `cmul` or its ampersand form `&c` of two Graßmann basis monomials can now be defined as follows: We have to split off recursively a single element from the first factor of the product and then use the Chevalley's Clifford map. Namely,

$$(\mathbf{e}_a \wedge \ldots \wedge \mathbf{e}_b \wedge \mathbf{e}_c) \,\&\mathtt{c}\, (\mathbf{e}_f \wedge \ldots \wedge \mathbf{e}_g) =$$
$$(\mathbf{e}_a \wedge \ldots \wedge \mathbf{e}_b) \,\&\mathtt{c}\, (\mathbf{e}_c \,\lrcorner_B (\mathbf{e}_f \wedge \ldots \wedge \mathbf{e}_g) + \mathbf{e}_c \wedge \mathbf{e}_f \wedge \ldots \wedge \mathbf{e}_g)$$
$$- ((\mathbf{e}_a \wedge \ldots \wedge \mathbf{e}_b) \llcorner_B \mathbf{e}_c) \,\&\mathtt{c}\, (\mathbf{e}_f \wedge \ldots \wedge \mathbf{e}_g). \quad (1)$$

Specifically, for $(\mathbf{e}_1 \wedge \mathbf{e}_2) \,\texttt{\&c}\, (\mathbf{e}_3 \wedge \mathbf{e}_4)$ we have

$$
\begin{aligned}
(\mathbf{e}_1 \wedge \mathbf{e}_2) \,\texttt{\&c}\, (\mathbf{e}_3 \wedge \mathbf{e}_4) &= (\mathbf{e}_1 \,\texttt{\&c}\, \mathbf{e}_2) \,\texttt{\&c}\, (\mathbf{e}_3 \wedge \mathbf{e}_4) - B(\mathbf{e}_1,\mathbf{e}_2)\mathbf{1} \,\texttt{\&c}\, (\mathbf{e}_3 \wedge \mathbf{e}_4) \\
&= \mathbf{e}_1 \,\texttt{\&c}\, (B(\mathbf{e}_2,\mathbf{e}_3)\mathbf{e}_4 - B(\mathbf{e}_2,\mathbf{e}_4)\mathbf{e}_3 + \mathbf{e}_2 \wedge \mathbf{e}_3 \wedge \mathbf{e}_4) \\
&\quad - B(\mathbf{e}_1,\mathbf{e}_2)\mathbf{1} \,\texttt{\&c}\, (\mathbf{e}_3 \wedge \mathbf{e}_4)
\end{aligned}
$$

and a second recursion of the process gives now

$$
\begin{aligned}
&= B(\mathbf{e}_2,\mathbf{e}_3)B(\mathbf{e}_1,\mathbf{e}_4) - B(\mathbf{e}_2,\mathbf{e}_4)B(\mathbf{e}_1,\mathbf{e}_3) + B(\mathbf{e}_2,\mathbf{e}_3)(\mathbf{e}_1 \wedge \mathbf{e}_4) \\
&\quad - B(\mathbf{e}_2,\mathbf{e}_4)(\mathbf{e}_1 \wedge \mathbf{e}_3) + \mathbf{B(e_1,e_2)(e_3 \wedge e_4)} - B(\mathbf{e}_1,\mathbf{e}_3)(\mathbf{e}_2 \wedge \mathbf{e}_4) \\
&\quad + B(\mathbf{e}_1,\mathbf{e}_4)(\mathbf{e}_2 \wedge \mathbf{e}_3) + \mathbf{e}_1 \wedge \mathbf{e}_2 \wedge \mathbf{e}_3 \wedge \mathbf{e}_4 - \mathbf{B(e_1,e_2)(e_3 \wedge e_4)}
\end{aligned}
$$

with the bolded terms canceling out. Note that the last term in the r.h.s. was superfluously generated in the first step of the recursion.

The Clifford product can be derived from the above recursion by linearity and associativity. The induction starts with a left factor of grade one or grade zero which is trivial, i.e., $\mathbf{1} \,\texttt{\&c}\, \mathbf{e}_a \wedge \ldots \wedge \mathbf{e}_b = \mathbf{e}_a \wedge \ldots \wedge \mathbf{e}_b$. In the case when the left factor is of grade one, we use the Clifford product expressed by the Clifford map of Chevalley, i.e., $\mathbf{e}_a \,\texttt{\&c}\, \mathbf{e}_b \wedge \ldots \wedge \mathbf{e}_c = \mathbf{e}_a \,\lrcorner_B (\mathbf{e}_b \wedge \ldots \wedge \mathbf{e}_c) + \mathbf{e}_a \wedge \mathbf{e}_b \wedge \ldots \wedge \mathbf{e}_c$. We make a complete induction in the following way: If the left factor is of higher grade, say $n$, one application of the recursion yields Clifford products where the new left factor is of grade either $n-1$ or $n-2$, hence the recursion stops after at most $n-1$ steps.

A disadvantage of the recursive approach is that additional terms are produced by shifting Graßmann wedge products into Clifford products in order to swap one factor to the right. While these terms eventually cancel out, their computation increases unnecessarily the total computing time. More importantly, they may easily exhaust any computer memory available and prevent Maple from completing the computation of the product.

An advantage of the recursive approach is realized when the bilinear form $B$ is numeric and sparse, that is, with many zeros. In this case, after each recursive step many terms drop out since Maple automatically simplifies expressions as soon as they are computed so that only a few remaining terms enter into the next step of the recursion. If the dimension of $V$ is large, i.e., $\dim V \geq 6$, computation with a sparse matrix $B$ benefits dramatically from this simplification process. It is then more efficient than the Hopf combinatorial algorithm encoded in `cmulRS` which computes all terms in the final output without taking advantage of the sparseness of $B$ except at the final evaluation step.

One could try to shift factors from the right to the left, however this works the same way and requires the same number of steps. Moreover, if the grade of the left factor $n$ is greater than the grade $m$ of the right factor, then the recursion stops also (since the terms evaluate to zero) after at most $n-1$ steps, so no increase in performance can be gained this way. For example, the above computation in Maple will be performed as follows:
```
> cmul(e1we2,e3we4);
```

$(B_{2,3} B_{1,4} - B_{2,4} B_{1,3}) \, Id + B_{2,3} \, e14 - B_{2,4} \, e13 - B_{1,3} \, e24 + B_{1,4} \, e23 + e1234$

Notice also that `cmul` accepts an arbitrary bilinear form $K$ as its argument:
```
> cmul[K](e1we2,e3we4);
```

$$(K_{2,3}\, K_{1,4} - K_{2,4}\, K_{1,3})\, Id + K_{2,3}\, e14 - K_{2,4}\, e13 - K_{1,3}\, e24 + K_{1,4}\, e23 + e1234$$

and likewise its ampersand form[8]
```
> &c[K](ei,ejwekwel);
```

$$eiwejwekwel + K_{i,j}\, ekwel - K_{i,k}\, ejwel + K_{i,l}\, ejwek$$

where we have also shown the ability of `CLIFFORD` to use symbolic indices. For clarity and to show our approach we display the algorithm of `cmulNUM` in Appendix A.

### 3.2 Procedure `cmulRS` based on the Rota-Stein combinatorial process

The procedure `cmulRS` is computed using the non-recursive Rota-Stein cliffordization. See [4, 5, 12, 20] and `BIGEBRA` help pages for additional references. The cliffordization process is based on the Hopf algebra theory. The Clifford product is obtained from the Graßmann wedge product and its Graßmann coproduct as shown by the following tangle:

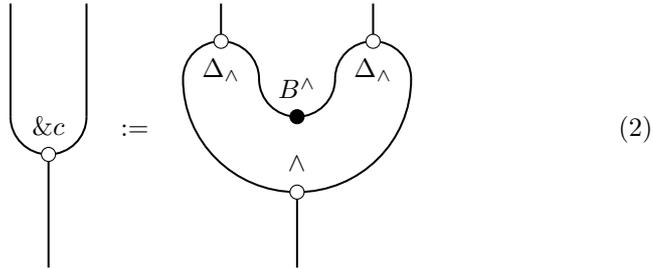

(2)

Here $\wedge$ is the Graßmann exterior wedge product and $\Delta_\wedge$ is the Graßmann exterior co-product which is obtained from the wedge product by a categorial duality: To every algebra over a linear space $A$ with a product we find a co-algebra with a co-product over the same space by reversing all arrows in all axiomatic commutative diagrams. Note that the co-product splits each input 'factor' $x$ into a sum of tensor products of ordered pairs $x_{(1)i}, x_{(2)i}$. The main requirement is that every such pair multiplies back to the input $x$ when the dual operation of multiplication is applied, i.e., $x_{(1)i} \wedge x_{(2)i} = x$ for each $i$-th pair. The 'cup' like part of the tangle decorated with $B^\wedge$ is the bilinear form $B$ on the generating space $V$ extended to the whole Graßmann algebra: It is a map $B^\wedge : \bigwedge V \times \bigwedge V \to k$ with $B : V \times V \to k$ evaluating to $B(\mathbf{x}, \mathbf{y})$ on vectors in $V$.

---

[8] Procedures `cmulNUM` and `cmulRS` do not have their special ampersand forms. Procedure `&c` uses internally `cmulNUM` or `cmulRS` depending on the current value of an environmental variable `_default_Clifford_product`. Current value of this and other environmental variables can be displayed by a procedure `CLIFFORD_ENV`.

Hence, `cmulRS` computes the Clifford product on Graßmann basis monomials $x$ and $y$ for the given $B$, which is later extended to Clifford polynomials by bilinearity, as follows:

$$\texttt{cmulRS}(x,y,B) = \sum_{i=1}^{n}\sum_{j=1}^{m}(\pm)x_{(1)i} \wedge y_{(2)j} B(x_{(2)i}, y_{(1)j}) \qquad (3)$$

where $n$ and $m$ give the cardinalities of the required splits and the sign is due to the parity of a permutation needed to arrange the factors.

A simplified algorithm of `cmulRS` looks as follows:

**cmulRS**(x,y,B) [x, y two Grassmann monomials, B - bilinear form]
**begin**
  lstx <- list of indices from x
  lsty <- list of indices from y
  NX <- length of lstx
  NY <- length of lsty
  funx <- function maps integers 1..NX onto elements of lstx keeping their order
  funy <- function maps integers 1..NY onto elements of lsty keeping their order
    (this is to calculate with arbitrary indices and to compute necessary signs)
  psetx <- power set of 1..NX (actually a list in a certain order)
    (the i-th and (2^NX+1-i)-th element are disjoint adding up to the set {1..NX})
  psety <- power set of 1..NY (actually a list in a certain order)
    (the i-th and (2^NY+1-i)-th element are disjoint adding up to the set {1..NY})
    (for faster computation we sort this power sets by grade)
    (we compute the sign for any term in the power set)
  psetx <- sort psetx by grade
  psety <- sort psety by grade
  pSgnx <- sum_(i in psetx) (-1)^sum_(j in psetx[i]) (psetx[i][j]-j)
  pSgny <- sum_(i in psety) (-1)^sum_(j in psety[i]) (psety[i][j]-j)
  (we need a subroutine for cup tangle computing the bilinear form cup(x,y,B))
    **begin cup**
      **if** |x| <> |y| **then** return 0 **end if**
      **if** |x| = 0 **then** return 1 **end if**
      **if** |x| = 1 **then** return B[x[1],y[1]] **end if**
      return sum_(j in 1..|x|)(-1)^(j-1)*B(x[1],y[j])*cup(x[2..-1],y/y[j],B)
    **end cup**
  (now we compute the double sum, to gain efficiency we do this grade wise)
  (note that there are r over NX r-vectors in psetx, analogously for psety)
  max_grade - |lstx <- convert_to_set union lsty <- convert_to_set|
  res <- 0, pos1 <- 0
  **for** j **from** 0 **to** NX (iterate over all j-vectors of psetx)
    **begin**
      F1 <- N1!/((N1-j)!*j!) (number of terms (N1 over j))
      pos2 <- 0
      **for** i **from** 0 **to** min(N2,max_grade-j)

```
    (iterate over all i-vectors of psety not exceeding max_grade while others are zero)
    begin
      F2 <- N2!/((N2-i)!*i!) (number of terms (N2 over i))
    for n from 1 to F1 (for all j-vectors)
      begin
        for m from 1 to F2 (for all i-vectors)
          begin
          res <- res + pSgnx[pos1+n]*pSgny[pos2+m]*
            *cup(fun1(psetx[PN1-pos1-n]),fun2(psety[pos2+m]),lname)*
            makeclibasmon -> ([fun1 -> psetx[pos1+n],fun2 -> psety[PN2-pos2-m]])
          end
      end
      pos2 <- pos2F2
    end
    pos1 <- pos1F1
  end
  reorder -> res (reorder basis elements in res into standard order)
end cmulRS
```

It is clear from this algorithm that only those terms are considered which might be non-zero: If all $B_{i,j}$ are non-zero and different so that no cancelation takes place between them, all these terms will survive. The combinatorial power of the Hopf algebraic approach is clearly demonstrated with this algorithm and its superior behavior shows up in benchmarks [4].

## 4  Dotted and undotted Graßmann bases in quantum Clifford algebras

### 4.1  On the meaning of the dotted wedge

We are going to show that Graßmann algebras with the dotted and the undotted wedge products are isomorphic. From a mathematical point of view, isomorphic objects are identical. The importance of studying otherwise isomorphic but not identical algebras comes from the fact that one is interested in various situations when one needs pairs of such algebras. This leads to a relative isomorphism, which is then mathematically and physically relevant. We just mention two places where the dotted wedge appears.

- In quantum field theory one needs to study various orderings of field operator products and/or correlation functions. In fermionic quantum field theory, a normal ordered product is expressed in terms of graded-commutative wedge products. A transition to time ordered products resp. correlation functions is equivalent to a transition to the dotted wedge products. The antisymmetric bilinear form in this case is called a *Wightman bilinear form*, see [9, 10, 11].

- In the theory of group representations one wants to deduce characters of subgroups of a given group by branching laws. If one derives the branching $U(n) \downarrow U(n-1)$ one encounters a pair of products which are related to the transition from the undotted to the dotted wedge, see [13].

In general, one can use Hopf algebra cohomology to classify maps which connect the various products. From this analysis it is known that algebra isomorphisms are related to 1-cocycles. The 1-cocycle condition guarantees that the transition is an algebra homomorphism. Below, we investigate in which way the wedge product –related to the creation operators– and the contraction –related to the annihilation operators– is affected by the algebra isomorphism induced by the antisymmetric part $F$ of a bilinear form $B$. This analysis can be extended to symmetric algebras [13] and to superspaces [9].

It was shown above that `CLIFFORD` uses the Graßmann algebra $\bigwedge V$ as the underlying vector space of the Clifford algebra $C\ell(V, B)$. Thus, the Graßmann wedge basis of monomials is the standard basis used in `CLIFFORD`. A general element $u$ in $C\ell(V, B)$ can be therefore viewed as a Graßmann polynomial.

When the bilinear form $B$ has an antisymmetric part $F = -F^T$, it is convenient to split it as $B = g + F$, where $g$ is the symmetric part of $B$, and to introduce the so called "dotted Graßmann basis" [6] and the dotted wedge product $\dot{\wedge}$. The original Graßmann basis will be referred to here as the "undotted Graßmann basis". In `CLIFFORD`, the wedge product is given by the procedure `wedge` and $\&w$ while the dotted wedge product is given by `dwedge` and $\&dw$.

According to the Chevalley definition of the Clifford product $\&c$, we have

$$\mathbf{x} \; \&\mathtt{c} \; u = \mathbf{x} \lrcorner_B u + \mathbf{x} \; \&\mathtt{w} \; u = \mathtt{LC}(\mathbf{x}, u, B) + \mathtt{wedge}(\mathbf{x}, u) \qquad (4)$$

for a 1-vector $\mathbf{x}$ and an arbitrary element $u$ of $C\ell(B)$. As before, $\mathtt{LC}(\mathbf{x}, u, B)$ denotes the left contraction of $u$ by $\mathbf{x}$ with respect to the bilinear form $B$. However, when $B = g + F$ then

$$\mathbf{x} \lrcorner_B u = \mathtt{LC}(\mathbf{x}, u, B) = \mathbf{x} \lrcorner_g u + \mathbf{x} \lrcorner_F u = \mathtt{LC}(\mathbf{x}, u, g) + \mathtt{LC}(\mathbf{x}, u, F) \qquad (5)$$

and

$$\begin{aligned} \mathbf{x} \; \&\mathtt{c} \; u &= \mathtt{LC}(\mathbf{x}, u, B) + \mathbf{x} \; \&\mathtt{w} \; u & (6) \\ &= \mathtt{LC}(\mathbf{x}, u, g) + \mathtt{LC}(\mathbf{x}, u, F) + \mathbf{x} \; \&\mathtt{w} \; u & (7) \\ &= \mathtt{LC}(\mathbf{x}, u, g) + \mathtt{dwedge}[F](\mathbf{x}, u) = \mathtt{LC}(\mathbf{x}, u, g) + \mathbf{x} \; \&\mathtt{dw} \; u & (8) \end{aligned}$$

where $\mathbf{x} \; \&\mathtt{dw} \; u = \mathbf{x} \; \&\mathtt{w} \; u + \mathtt{LC}(\mathbf{x}, u, F)$. That is, the wedge and the dotted wedge "differ" by the contraction term(s) with respect to the antisymmetric part $F$ of $B$. This dotted wedge $\&dw$ can be extended to elements of higher grades. Its properties are discussed next.

## 4.2 Indexing `dwedge` and $\&dw$

Procedure `dwedge` (and its infix form $\&dw$) requires an index which can be a symbol or an antisymmetric matrix. That is, `dwedge` computes the dotted wedge product of two Graßmann polynomials and expresses its answer in the undotted basis. Special procedures exist which convert polynomials between the undotted and dotted bases. When no index is used, the default is $F$:
```
> dwedge[K](e1+2*e2we3,e4+3*e1we2);&dw(ei+2*ejwek,ei+2*ejwek);
```

$$-(-K_{1,4} + 6\,K_{2,3}\,K_{1,2})\,Id - 6\,K_{1,2}\,e2we3 - 6\,K_{2,3}\,e1we2 - 2\,K_{2,4}\,e3 +$$
$$2\,K_{3,4}\,e2 - 3\,K_{1,2}\,e1 + e1we4 + 2\,e2we3we4$$
$$4\,eiwejwek - 4\,F_{i,k}\,ej + 4\,F_{i,j}\,ek - 8\,F_{j,k}\,ejwek - 4\,F_{j,k}{}^2\,Id$$

Observe that conversion from the undotted wedge basis to the dotted wedge basis using antisymmetric form $F$ and dwedge[F] are related through the following convert function:

$$\mathtt{dwedge[F]}(e1, e2, ..., en) = \mathtt{convert}(e1we2w...wen, \mathtt{wedge\_to\_dwedge}, F)$$

which can be shown as follows:
```
> F:=array(1..9,1..9,antisymmetric):
> dwedge[F](e1,e2)=convert(wedge(e1,e2),wedge_to_dwedge,F);
```

$$e1we2 + F_{1,2}\,Id = e1we2 + F_{1,2}\,Id$$

```
> dwedge[F](e1,e2,e3)=convert(wedge(e1,e2,e3),wedge_to_dwedge,F);
```

$$e1we2we3 + F_{2,3}\,e1 - F_{1,3}\,e2 + F_{1,2}\,e3 = e1we2we3 + F_{2,3}\,e1 - F_{1,3}\,e2 + F_{1,2}\,e3$$

$$C\ell(B)_\wedge \xrightleftharpoons[\mathtt{dwedge\_to\_wedge}]{\mathtt{wedge\_to\_dwedge}} C\ell(B)_{\dot\wedge}$$

Diagram 1: Isomorphisms between $C\ell(B)_\wedge$ and $C\ell(B)_{\dot\wedge}$.

### 4.3 Associativity of dwedge

Operation dwedge is associative with the unity $\mathbf{1} = \mathtt{Id}$ as a unit:
```
> evalb(dwedge[F](dwedge[F](e1,e2),e3)=dwedge[F](e1,dwedge[F](e2,e3)));
```

*true*

Associativity can be encoded in a commutative diagram, see diagram 2. It was checked with CLIFFORD up to dimension 5.[9]
For some arbitrary random Clifford polynomials[10] $u, v, z$ expressed in Graßmann undotted basis we can show associativity as follows:

---

[9]There is a proof making use of Hopf algebra cohomology that any 2-cocycle deformation comes up with an associative product. The deformation by $F$ is such a 2-cocycle and associativity can be proved for the general case.

[10]In CLIFFORD ver. 6 and higher there are three procedures useful for testing that return a random Graßmann basis monomial, a random monomial and a random polynomial, respectively. See ?rd_clibasmon, ?rd_climon, ?rd_clipolynom.

$$\begin{array}{ccc}
C\ell(B)_{\dot{\wedge}} \otimes C\ell(B)_{\dot{\wedge}} \otimes C\ell(B)_{\dot{\wedge}} & \xrightarrow{\texttt{dwedge[F]} \otimes \mathbf{1}} & C\ell(B)_{\dot{\wedge}} \otimes C\ell(B)_{\dot{\wedge}} \\
{\scriptstyle \mathbf{1} \otimes \texttt{dwedge[F]}} \downarrow & & \downarrow {\scriptstyle \texttt{dwedge[F]}} \\
C\ell(B)_{\dot{\wedge}} \otimes C\ell(B)_{\dot{\wedge}} & \xrightarrow{\texttt{dwedge[F]}} & C\ell(B)_{\dot{\wedge}}
\end{array}$$

Diagram 2: Associativity of `dwedge[F]` in $C\ell(B)_{\dot{\wedge}}$.

```
> u:=2*Id+e1-3*e2we3:v:=3*Id-4*e1we3+e7:z:=4*Id-2*e3+e1we2we3:
> evalb(dwedge[F](Id,u)=u),evalb(dwedge[F](u,Id)=u);
```

$$true, true$$

```
> evalb(dwedge[F](dwedge[F](u,v),z)=dwedge[F](u,dwedge[F](v,z)));
```

$$true$$

We have, therefore, the following identity that expresses an isomorphism between two Graßmann algebras: dotted and undotted. For any two elements $u$ and $v$ in $C\ell(B)$, $B = g + F$, that are, by default, expressed in terms of the undotted Graßmann basis, we find:

$$u \wedge v = (u_F \dot{\wedge} v_F)_{-F}. \tag{9}$$

Here $u_F$ and $v_F$ are the elements $u$ and $v$ expressed in the dotted basis with respect to the form $F$ while $(\ldots)_{-F}$ denotes conversion back from the dotted basis to the undotted basis w.r.t. $-F = F^T$. $C\ell(B)_{\wedge}$ and $C\ell(B)_{\dot{\wedge}}$ denote the modules w.r.t. the two filtrations in use.

$$\begin{array}{ccc}
C\ell(B)_{\wedge} \otimes C\ell(B)_{\wedge} & \xrightarrow{(\ldots)_F \otimes (\ldots)_F} & C\ell(B)_{\dot{\wedge}} \otimes C\ell(B)_{\dot{\wedge}} \\
{\scriptstyle \wedge} \downarrow & & \downarrow {\scriptstyle \dot{\wedge}} \\
C\ell(B)_{\wedge} & \xleftarrow{(\ldots)_{-F}} & C\ell(B)_{\dot{\wedge}}
\end{array}$$

Diagram 3: Relation between $\wedge$ and $\dot{\wedge}$ products.

This can be illustrated in `CLIFFORD` as follows:
```
> uu:=convert(u,wedge_to_dwedge,F); vv:=convert(v,wedge_to_dwedge,F);
```

$$uu := e1 - 3\,e2we3 - 3\,F_{2,3}\,Id + 2\,Id$$

$$vv := 3\,Id - 4\,e1we3 - 4\,F_{1,3}\,Id + e7$$

```
> out1:=dwedge[F](uu,vv):  #dwedge computed w.r.t.  F
> out2:=convert(out1,dwedge_to_wedge,-F); #back to undotted basis
```

$$out2 := 3\,e1 - 9\,e2we3 + 6\,Id - 8\,e1we3 + e1we7 - 3\,e2we3we7 + 2\,e7$$

```
> out3:=wedge(u,v); #direct computation of wedge product
```

$$out3 := 3\,e1 - 9\,e2we3 + 6\,Id - 8\,e1we3 + e1we7 - 3\,e2we3we7 + 2\,e7$$

and it can be seen that out2 = out3 establishing the relation:

## 4.4 Dotted and undotted wedge bases

The default Graßmann basis in $C\ell(B)$ used in CLIFFORD is undotted. However, one can easily use the dotted basis. For example, we expand the basis of the original wedge into the dotted wedge, and back. For this purpose we choose $\dim V = 3$ and consider $C\ell(B)$ with the antisymmetric part $F$. The undotted wedge basis for $\bigwedge V$ is then:
```
> w_bas:=cbasis(dim_V); #the wedge basis
```

$$w\_bas := [Id,\, e1,\, e2,\, e3,\, e1we2,\, e1we3,\, e2we3,\, e1we2we3]$$

Now we map the convert function onto this basis to get the dotted wedge basis:
```
> d_bas:=map(convert,w_bas,wedge_to_dwedge,F);
> test_wbas:=map(convert,d_bas,dwedge_to_wedge,-F);
```

$$d\_bas := [Id,\, e1,\, e2,\, e3,\, e1we2 + F_{1,2}\,Id,\, e1we3 + F_{1,3}\,Id,\, e2we3 + F_{2,3}\,Id,$$
$$e1we2we3 + F_{2,3}\,e1 - F_{1,3}\,e2 + F_{1,2}\,e3]$$

$$test\_wbas := [Id,\, e1,\, e2,\, e3,\, e1we2,\, e1we3,\, e2we3,\, e1we2we3]$$

Notice that only the unity **1** and the one vector basis elements $\mathbf{e}_i$ remain unaltered and that the other basis elements of higher grades pick up additional terms of lower grades (which preserves the filtration). It is possible to define aliases in CLIFFORD for the dotted wedge basis "monomials" similar to the Graßmann basis monomials. For example, we could denote the element $e1we2 + F[1,2]*Id$ by $e1We2 (= \mathbf{e}_1 \dot\wedge \mathbf{e}_2)$ and similarly for other elements:
```
> alias(e1We2=e1we2 + F[1,2]*Id,e1We3=e1we3 + F[1,3]*Id,
> e2We3=e2we3 + F[2,3]*Id,
> e1We2We3=e1we2we3+F[2,3]*e1-F[1,3]*e2+F[1,2]*e3);
```

$$I,\, e1We2,\, e1We3,\, e2We3,\, e1We2We3$$

and then Maple will automatically display dotted basis in terms of the aliases:
```
> d_bas;
```

$$[Id,\, e1,\, e2,\, e3,\, e1We2,\, e1We3,\, e2We3,\, e1We2We3]$$

That is, as linear spaces we find the isomorphism:

$$\begin{aligned} C\ell(B) &\cong \langle \mathbf{1}, \mathbf{e}_1, \mathbf{e}_2, \mathbf{e}_3, \mathbf{e}_1 \wedge \mathbf{e}_2, \mathbf{e}_1 \wedge \mathbf{e}_3, \mathbf{e}_2 \wedge \mathbf{e}_3, \mathbf{e}_1 \wedge \mathbf{e}_2 \wedge \mathbf{e}_3 \rangle \\ &\cong \langle \mathbf{1}, \mathbf{e}_1, \mathbf{e}_2, \mathbf{e}_3, \mathbf{e}_1 \dot\wedge \mathbf{e}_2, \mathbf{e}_1 \dot\wedge \mathbf{e}_3, \mathbf{e}_2 \dot\wedge \mathbf{e}_3, \mathbf{e}_1 \dot\wedge \mathbf{e}_2 \dot\wedge \mathbf{e}_3 \rangle \end{aligned}$$

where $\mathbf{e}_1 \dot\wedge \mathbf{e}_2 = e1We2$, etc.

### 4.5 Contraction in dotted and undotted bases

The contraction $\lrcorner_K$ w.r.t. any bilinear form $K$ works on both undotted and dotted bases in the same manner which can be seen if we re-convert the dotted basis after the computation into the wedge (undotted) basis. In a reasonable setting, the antisymmetric bilinear form $F$ would be the antisymmetric part of $B$. To read more about the left contraction LC in $C\ell(B)$ check the help page for LC or see [6]. We have the following identity for any two elements $u$ and $v$ in $C\ell(B)$ expressed in the undotted Graßmann basis:

$$v \lrcorner_B u = (v \lrcorner_B u_F)_{-F} \tag{10}$$

$$\begin{array}{ccc} C\ell(B)_\wedge \otimes C\ell(B)_\wedge & \xrightarrow{\mathbf{1} \otimes (\ldots)_F} & C\ell(B)_\wedge \otimes C\ell(B)_{\dot\wedge} \\ \downarrow {\lrcorner_B} & & \downarrow {\lrcorner_B} \\ C\ell(B)_\wedge & \xleftarrow{(\ldots)_{-F}} & C\ell(B)_{\dot\wedge} \end{array}$$

Diagram 4: Contraction w.r.t. wedge and dotted wedge.

As before, $u_F$ is the element $u$ expressed in the dotted basis while $(\ldots)_{-F}$ accomplishes conversion back to the undotted basis. To illustrate this fact, we first contract from the left an arbitrary element $u$ in $C\ell(B)$ by $\mathbf{1}, \mathbf{e}_i, \mathbf{e}_i \wedge \mathbf{e}_j, \mathbf{e}_i \wedge \mathbf{e}_j \wedge \mathbf{e}_k$, $1 \leq i,j,k \leq 3$ (here we limit our example to $\dim V = 3$) and then we extend it to a left contraction by an arbitrary element $v$ in $C\ell(B)$.

```
> u:=add(x.i*w_bas[i+1],i=0..7):uF:=convert(uw,wedge_to_dwedge,F):
> v:=add(y.i*w_bas[i+1],i=0..7):
```
Contraction with respect to $\mathbf{1}$:
```
> evalb(LC(Id,u,B)=convert(LC(Id,uF,B),dwedge_to_wedge,-F));
```

$$true$$

Contraction with respect to $\mathbf{e}_i$:
```
> evalb(LC(ei,u,B)=convert(LC(ei,uF,B),dwedge_to_wedge,-F));
```

$$true$$

Contraction with respect to $\mathbf{e}_i \wedge \mathbf{e}_j$:
```
> evalb(LC(eiwej,u,B)=convert(LC(eiwej,uF,B),dwedge_to_wedge,-F));
```

$$true$$

Contraction with respect to $\mathbf{e}_i \wedge \mathbf{e}_j \wedge \mathbf{e}_k$ :
```
> evalb(LC(eiwejwek,u,B)=convert(LC(eiwejwek,uF,B),dwedge_to_wedge,-F));
```

$$true$$

Finally, contraction with respect to an arbitrary element $v$ :
```
> evalb(LC(v,u,B)=convert(LC(v,uF,B),dwedge_to_wedge,-F));
```

$$true$$

## 4.6 Clifford product in dotted and undotted bases

We can build a Clifford algebra $C\ell(B)$ over each basis set, that is, over the undotted or dotted Graßmann basis, but with different bilinear forms: $B = g$ or $B = g + F$ respectively (following notation from [6]). Let us compute various Clifford products with respect to the symmetric form $g$ and with respect to the full form $B$ using procedure `cmul` that takes a bilinear form as its index. As an example, we will use two most general elements $u$ and $v$ in $\bigwedge V$ when $\dim V = 3$. Most output will be eliminated.
```
> u:=add(x.k*w_bas[k+1],k=0..7):v:=add(y.k*w_bas[k+1],k=0..7):
```
We can then define in $\bigwedge V$ a Clifford product `cmul[g]` with respect to the symmetric part $g$ and another Clifford product `cmul[B]` with respect to the entire form $B$ :
```
> cmulg:=proc() return cmul[g](args) end proc:
> cmulB:=proc() return cmul[B](args) end proc:
```
Thus, we are ready to perform computations around our next commutative diagram, however most output will be eliminated to save space.

$$\begin{array}{ccc}
C\ell(g)_{\wedge} \otimes C\ell(g)_{\wedge} & \xrightarrow{(\ldots)_F \otimes (\ldots)_F} & C\ell(g)_{\dot{\wedge}} \otimes C\ell(g)_{\dot{\wedge}} \\
\downarrow \mathtt{cmul[g]} & & \downarrow \mathtt{cmul[B]} \\
C\ell(g)_{\wedge} & \xleftarrow{(\ldots)_{-F}} & C\ell(g)_{\dot{\wedge}}
\end{array}$$

Diagram 5: Clifford multiplications `cmul[g]` and `cmul[B]` w.r.t. dotted and undotted basis.

First, we compute the Clifford product `cmul[g]`$(u,v)$ in $C\ell(g)$ in undotted Graßmann basis.
```
> uv:=cmulg(u,v):  #Clifford product w.r.t.  g in Cl(g) in wedge basis
```
Now, we convert $u$ and $v$ to $u_F$ and $v_F$, respectively, expressed in the dotted wedge basis:
```
> uF:=convert(u,wedge_to_dwedge,F):vF:=convert(v,wedge_to_dwedge,F):
```
We now compute the Clifford product of $u_F$ and $v_F$ in $C\ell(B)$ in the dotted wedge basis,
```
> uFvF:=cmulB(uF,vF): #Clifford product in Cl(B) in dwedge basis
```
convert back the above result back to the undotted wedge basis:
```
> uv2:=convert(uFvF,dwedge_to_wedge,-F): #convert result dwedge->wedge
```
and verify that the results are the same:

```
> simplify(uv-uv2); #show equality!
```

$$0$$

Thus, we have shown that the following identity involving `cmul[g]` and `cmul[B]` is true (at least when $\dim V = 3$).[11] The result is folklore, and may be found e.g. in [7, 15].

$$(u\,v)_g = u\,\&\mathtt{c}_g\,v = (u_F\,\&\mathtt{c}_B\,v_F)_{-F} = ((u_F\,v_F)_B)_{-F} \tag{11}$$

This shows that the Clifford algebra $C\ell(g)$ of the symmetric part $g$ of $B$ using the undotted exterior basis is isomorphic, as an associative algebra, to the Clifford algebra $C\ell(B)$ of the entire bilinear form $B = g + F$ spanned by the dotted wedge basis if the antisymmetric part $F$ of $B$ is exactly the same as $F$ used to connect the two bases.

$$(\ldots)_F \in \mathrm{Hom}_{\mathrm{Alg}}(C\ell(g), C\ell(B)), \qquad B = g + F$$

### 4.7 Reversion in dotted and undotted bases

We proceed to show that the expansion of the Clifford basis elements into the dotted or undotted exterior products has also implications for other well known operations such as the Clifford reversion anti-automorphism $\tilde{\ }: C\ell(B) \to C\ell(B)$, $uv \mapsto \tilde{v}\tilde{u}$, which preserves the grades in $\dot{\bigwedge} V$ [but not in $\bigwedge V$ unless $B$ is symmetric.] Only when the bilinear form is symmetric, we find that the reversion is grade preserving, otherwise it reflects only the filtration: That is, reversed elements are in general sums of terms of the same and lower degrees.

```
> reversion(e1we2,B); #reversion with respect to B
> reversion(e1we2,g); #reversion with respect to g (classical result)
```

$$-e1we2 - 2\,F_{1,2}\,Id$$

$$-e1we2$$

To provide a clear idea how the various reversions are related, we consider the following commuting diagram:

$$\begin{array}{ccc} C\ell(B)_\wedge & \xrightarrow{\mathtt{reversion}[B](\ldots)} & C\ell(B)_{\dot\wedge} \\ {\scriptstyle \mathtt{reversion}[g](\ldots)}\Big\downarrow & \nearrow {\scriptstyle (\ldots)_{2F}} & \\ C\ell(B)_\wedge & & \end{array}$$

**Diagram 6:** Relation between `reversion[g]` and `reversion[B]` and the basis transformation $(\ldots)_{2F}$.

The reader should note that the map, depicted by the diagonal arrow, involves a change of basis induced by the antisymmetric bilinear form $2\,F$ and not $F$. The factor 2 is crucial and appears due to an asymmetry between the undotted and dotted bases.

---

[11] Here, $(u\,v)_g$ is the Clifford product with respect to $g$ while $u_F\,\&\mathtt{c}_B\,v_F$ and $(u_F v_F)_B$ are the Clifford products with respect to $B$, that is, in $C\ell(g)$ and $C\ell(B)$, respectively.

This suggests to introduce a symmetrically related *triple* of bases w.r.t. $-F/2$, $F \equiv 0$ and $F/2$. In such a setting, $F$ (resp. $-F$) connects the two dotted bases induced by $\pm F/2$.

Observe in the pre-last display above that only when $B_{1,2} = B_{2,1}$, the result $-\mathbf{e}_1 \wedge \mathbf{e}_2$ known from the theory of classical Clifford algebras is obtained. Likewise,
> `cbas:=cbasis(3);`

$$cbas := [Id,\ e1,\ e2,\ e3,\ e1we2,\ e1we3,\ e2we3,\ e1we2we3]$$

> `map(reversion,cbas,B);`

$[Id,\ e1,\ e2,\ e3,\ -e1we2 - 2\,F_{1,2}\,Id,\ -e1we3 - 2\,F_{1,3}\,Id,\ -e2we3 - 2\,F_{2,3}\,Id,$
$-2\,F_{2,3}\,e1 + 2\,F_{1,3}\,e2 - 2\,F_{1,2}\,e3 - e1we2we3]$

If instead of $B$ we use a symmetric matrix $g = g^T$ (or the symmetric part of $B$), then
> `map(reversion,cbas,g);`

$$[Id,\ e1,\ e2,\ e3,\ -e1we2,\ -e1we3,\ -e2we3,\ -e1we2we3]$$

Convert now $\mathbf{e}_1 \wedge \mathbf{e}_2$ to the dotted basis to get $\mathbf{e}_1 \dot\wedge \mathbf{e}_2 = e1We2$:
> `convert(e1we2,wedge_to_dwedge,F);`

$$e1We2$$

Apply reversion to $e1We2$ with respect to $F$ to get the reversed element in the dotted basis:
> `reversed_e1We2:=reversion(e1We2,F);`

$$reversed\_e1We2 := -e1we2 - F_{1,2}\,Id$$

Observe, that the above element is equal to the negative of $e1We2$ just like reversing $e1we2$ with respect to the symmetric part $g$ of $B$:
> `reversed_e1We2+e1We2;`

$$0$$

Finally, convert reversed $e1We2$ to the undotted standard Graßmann basis to get $-e1we2$:
> `convert(reversed_e1We2,dwedge_to_wedge,-F);`

$$-e1we2$$

The above, of course, can be obtained by applying reversion to $e1we2$ with respect to the symmetric part $g$ of $B$:
> `reversion(e1we2,g); #reversion w.r.t. the symmetric part g`

$$-e1we2$$

This shows that the dotted wedge basis is the particular basis which is stable under the Clifford reversion computed with respect to $F$, the antisymmetric part of the bilinear form $B$. This requirement allows one to distinguish Clifford algebras $C\ell(g)$ which have a symmetric bilinear form $g$ from those which do not have such symmetric bilinear form but a more general form $B$ instead. We call the former **classical Clifford algebras** while we use the term **quantum Clifford algebras** for the general not necessarily symmetric case [2].

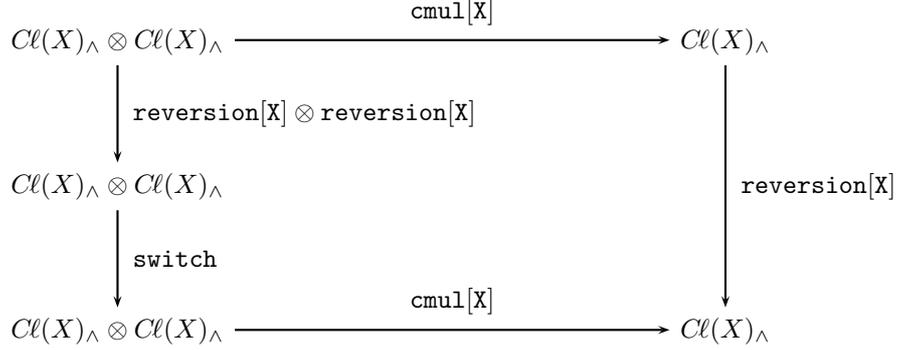

Diagram 7: Relation between the reversion[X] of type $X \in \{g,F,B\}$ with the corresponding Clifford multiplication cmul[X]. The map called switch is the ungraded switch of tensor factors, that is, $\text{switch}(A \otimes B) = B \otimes A$.

## 5 Conclusions

This paper continues with the second part [5] about BIGEBRA where further aims and outlooks for the future applications of CLIFFORD and BIGEBRA are given.

Appendix A: Code of cmulNUM

Here is a shortened code of the recursive procedure cmulNUM.

**cmulNUM**(a1,a2,B) [a1, a2 - two Grassmann monomials, B - name of bilinear form]
**begin**
  **if** nargs <>3 **then error** "exactly three arguments are needed" **end if**
  **if** has(0,map(simplify,[a1,a2])) **then return** 0 **end if**
  **if** a2='Id' **then return** a1 **end if**
  **if** a1='Id' **then return** a2 **end if**
L <- indices from a1
N <- length of L
coB,nameB <- coefficient of B, B [to handle -B]
  **if** N=0 **then return** coeff(a1,Id)*a2 **elif** N=1 **then**
L2 <- list of indices from a2
  **return** reorder(simplify(makeclibasmon([L[1],op(L2)])
    +add((-1)^(i-1)*coB*nameB[L[1],L2[i]]*
    makeclibasmon(subs(L2[i]=NULL,L2)),i=1..nops(L2))))
  **elif** N=2 **then**
    x1 <- substring(a1,1..2)

```
      x2 <- substring(a1,4..5)
      p2 <- procname(x2,a2,B)
      S <- clibilinear(x1,p2,procname,B)
      return simplify(S-coB*nameB[op(L)]*a2)
   end if;
x <- cat(e,L[-1])
p1 <- substring(a1,1..(3*N-4))
p2 <- procname(x,a2,B)
S <- clibilinear(p1,p2,procname,B)
   -add((-1)^(i)*coB*nameB[L[-i],L[-1]]*
   procname(makeclibasmon(subs(L[-i]=NULL,L[1..-2])),a2,B),i=2..N)
return reorder(simplify(S))
end cmulNUM
```

**Acknowledgment:**

The first author, R.A., gratefully acknowledges financial support from the College of Arts and Sciences, Tennessee Technological University, to present a preliminary version of this paper at the ACA 2002 in Volos, Greece.